\documentclass[
    ,final            
  ,draft            
  ,numberedheadings 
  square,
  sort,
  comma,
  numbers
  ]
  {aipproc}

\usepackage[mathscr]{eucal}
 \usepackage{amsmath, amssymb, bm}

   \def\degr{\hbox{$^\circ$}}
\newcommand{\be}{\begin{equation}}
\newcommand{\ee}{\end{equation}}
\newcommand{\bdm}{\begin{displaymath}}
\newcommand{\edm}{\end{displaymath}}
\layoutstyle{6x9}

\renewcommand{\vec}[1]{{\bm{#1}}}
\newcommand{\fr}[2]{{\displaystyle \frac{#1}{#2}}}
\newcommand{\sfr}[2]{{{#1}/{#2}}}

\newcommand{\pdiff}[2]{{\fr{\partial{#1}}{\partial{#2}}}}

\newcommand{\aver}[1]{{\langle #1 \rangle}}
\newcommand{\rotor}{\mathop{\rm rot}\nolimits}

\def\degr{\hbox{$^\circ$}}

\begin{document}
\noindent {\it To appear in Astronomy Reports, 2016, Vol. 60, No. 5}

\vspace{0.2 cm}

\title{Features of the Matter Flows
in the Peculiar Cataclysmic Variable AE Aquarii}


\author{P.B.\,Isakova}
{address={Institute of Astronomy, Russian Academy of Sciences, ul.
Pyatnitskaya 48, Moscow, 119017 Russia},,email={isakovapb@inasan.ru} }
\author{N.R.\,Ikhsanov} {address={Central
(Pulkovo) Astronomical Observatory, Russian Academy of Sciences, Pulkovskoe
shosse 65, St. Petersburg, 196140 Russia}, altaddress= {St. Petersburg State
University, Universitetskii pr. 28, Petrodvorets, 198504 Russia}}
\author{A.G.\,Zhilkin}
{address={Institute of Astronomy, Russian Academy of Sciences, ul.
Pyatnitskaya 48, Moscow, 119017 Russia}, altaddress= {Chelyabinsk State
University, ul. Bratyev Kashirinykh 129, Chelyabinsk, 454001 Russia}}
\author{D.V.\,Bisikalo}
{address={Institute of Astronomy, Russian Academy of Sciences, ul.
Pyatnitskaya 48, Moscow, 119017 Russia}}
\author{N.G.\,Beskrovnaya}
{address={Central (Pulkovo) Astronomical Observatory, Russian Academy of
Sciences, Pulkovskoe shosse 65, St. Petersburg, 196140 Russia}}

\begin{abstract}

The structure of plasma flows in close binary systems in which one of the
components is a rapidly rotating magnetic white dwarf is studied. The main
example considered is the AE~Aquarii system; the spin period of the
 white dwarf is about a factor of 1000 shorter than the orbital
period, and the magnetic field on the white dwarf surface is of order of 50
MG. The mass transfer in this system was analyzed via numerical solution of
the system of MHD equations. These computations show that the magnetic field
of the white dwarf  does not significantly influence the velocity field of the
material in its Roche lobe in the case of laminar flow regime, so that the
field does not hinder the formation of a transient disk (ring) surrounding the
magnetosphere. However, the efficiency of the energy and angular momentum
exchange between the white dwarf and the surrounding material increases
considerably with the development of turbulent motions in the matter,
resulting in its acceleration  at the magnetospheric boundary and further
 escape from the system at a high rate. The time scales of the
transition of the system between the laminar and turbulent modes are close to
those of the AE\,Aqr transition  between its quiet and active phases.

\end{abstract}

\maketitle


\section{Introduction}

AE~Aquarii is one of the most actively studied cataclysmic variables
\cite{Warner}. It is located at a relatively small distance ($d = 100 \pm
30$~pc) and has a binary orbital period, $P_\text{orb} \simeq 9.88$~hr, and a
low eccentricity, $e \simeq 0.02$ \cite{Welsh-etal-1995, friedjung97}. The
binary contains a K3--K5 red dwarf and a strongly magnetized white dwarf which
rotates at extremely short period, $P_\text{spin} \simeq 33$~s
\cite{Patterson-1979}. The limits for the orbital inclination are $43^{\degr}
< i < 70^{\degr}$, with the most probable value of $55^{\degr} \pm 7^{\degr}$.
The ratio of the masses of the red dwarf, $M_\text{d}$, and the white dwarf,
$M_\text{a}$, is $q = M_\text{d}/M_\text{a} \sim 0.6$--$0.8$ (see
\cite{Reinsch-Beuermann-1994, Welsh-etal-1995, Ikhsanov2012} and references
therein).

The bolometric luminosity of the system from the radio to the X-rays is
$L_{\rm bol} \simeq 10^{33}~\text{erg/s}$. A relatively small fraction
($10^{-6}$--$10^{-4}$) of this energy is emitted in the radio, where the
system manifests itself as a non-thermal variable source
\cite{Bastian-etal-1988}. The radiation is predominantly thermal in other
wavebands, and can be well approximated by a superposition of three spatially
separated sources. The radiation of the red dwarf dominates in the infrared
and optical domains (70\%--95\%). The contribution of the white dwarf does not
exceed 2\% of the bolometric luminosity of the system; the white dwarf is
revealed through its regular pulsations with the period of 33~s, which are
observed in the optical, ultraviolet (UV), and X-ray. The third source has a
strongly variable continuum and emission lines in the optical, UV, and X-rays.
This source is associated with the material which moves in the Roche lobe of
the white dwarf and flows away from the binary. This source is probably
responsible for the unique flaring activity of the star, which has no analogs
among all known classes of non-stationary objects.

Henize \cite{Henize-1949} was the first to report the unusually rapid
variations of AE~Aquarii; he remarked that the flares of the star occur one
after another, with a recurrence time of about an hour. Further observations
have shown that the star predominantly stays in its active phase: the total
duration of observations in the active state (for observations covering more
than 65~years) exceeded the total duration of observations in the quiet state.
Five-to-ten-minute flares are observed in the active phase, they overlap,
forming series with durations from several tens of minutes to several hours.
The brightness  variations during flares occur synchronously in the optical,
UV, and Xrays. The maximum energy release over the entire flare is in the UV,
where the luminosity of the system can change by an order of magnitude over
several minutes. The transition from the active to the quiet state happens
without any apparent intermediate phase \cite{Beskrovnaya-etal-1996}. In the
quiet state, whose duration is usually several tens of minutes, reaching
two-to-three hours only in rare cases, the radiation of the system is
characterized by the presence of flickering \cite{Bruch-1991,
Eracleous-Horne-1996}.

The temporal structure of the flares observed in the
radio is similar. However, the variations of the radio
intensity do not correlate with those in other parts
of the spectrum. VLBI observations of AE~Aquarii
show that the size of the radio source is apparently
comparable to the size of the system, and that the
radio emission is obviously non-thermal \cite{de-Jager-1994}. This
suggests the action of different mechanisms generating
the flares in the radio and optical, while the trigger
mechanism responsible for the time characteristics of
these events is probably the same for the system as a
whole \cite{Abada-Simon-etal-2005}.

The most probable reason for the unique flaring activity of AE~Aquarii is an
unusual state of its white dwarf. The observed spindown rate of
the object, $\dot{P}_0 = (5.64 \pm 0.02) \times 10^{-14}$~s/s
\cite{de-Jager-etal-1994, Welsh-1999}, implies that its spindown power,
\begin{equation}\label{eq1.1}
 L_{\rm sd} = I \omega_{\rm s} \dot{\omega}_{\rm s} \simeq
 6 \times 10^{33} I_{50} P_{33}^{-3}
 \left( \frac{\dot{P_{\rm s}}}{\dot{P}_0} \right)~\text{erg}\cdot\text{s}^{-1},
\end{equation}
is a factor of 120--300 higher than the luminosity of the system in the UV and
X-rays, and more than a factor of five higher than its bolometric luminosity,
$L_{\rm bol}$. Here,  $\omega_{\rm s} = 2 \pi/P_\text{spin}$ is the angular
velocity of the white dwarf rotation; $\dot{\omega}_{\rm s} = d\omega_{\rm
s}/dt$; $P_{33}$ and $I_{50}$ are the period and the moment of inertia of the
white dwarf  in units of 33\,s and $10^{50}~\text{g}\cdot\text{cm}^2$, and
$\dot{P_{\rm s}} = dP_{\rm spin}/dt$. This situation, with the spindown power
of the star  considerably exceeding its luminosity, is quite atypical for
cataclysmic variables. Among all types of sources in our Galaxy, this property
is characteristic only of ejecting pulsars, whose spindown power is
transformed predominantly into the energy of relativistic winds and
electromagnetic and MHD waves. Ikhsanov \cite{Ikhsanov-1998} was the first to
draw attention to this situation, having estimated the magnetic moment of the
white dwarf using the formula for magneto-dipole losses:
\begin{equation}\label{magmom}
 \mu \simeq 10^{34} P_{33}^{2}
 \left(
 \frac{L_{\rm sd}}{6 \times 10^{33}~\text{erg}\cdot\text{s}^{-1}}
 \right)^{1/2}~\text{G}\cdot\text{cm}^3.
\end{equation}
In this case, the mean magnetic field at the surface of
the white dwarf is $\sim 50$~MG (see also \cite{Ikhsanov-Biermann-2006}).

The basis for the first doubts about the accretion
nature of the emission from AE~Aquarii was provided
by observations with the Hubble Space Telescope
\cite{Eracleous-etal-1994}. They made it possible to identify the
source of the 33s (and 16.5~s) pulsations observed
in the optical and UV with two diametrically opposite spots located on the
surface of the white dwarf
(probably in the regions of its magnetic poles). The
temperature of these regions, $T_{\rm p} \sim 26\,000$~K,
exceeds the mean surface temperature of the white
dwarf, $T_{\rm wd} \sim 10\,000$--$16\,000$~K, by no more than a
factor of three. It was also found that the intensity
of the pulsating component in the optical and UV
remained virtually constant during flares, ruling out
explanations of the flaring activity of the system in terms of
non-stationary accretion of material onto the white dwarf
surface.

These doubts were made stronger by the results
of X-ray observations. The X-ray spectrum, with a
much lower intensity than in the UV, was found to be
exceptionally soft \cite{Osborne-etal-1995, Choi-etal-1999}, resembling a coronal spectrum
rather than an accretion-powered spectrum \cite{de-Jager-1994}.
Moreover, the source of non-pulsating X-rays, whose
contribution exceeds 80\% in quiescence and reaches
93\% during flares \cite{Choi-Dotani-2006}, is larger in size than the white dwarf
radius by  a factor of more than ten \cite{Itoh-etal-2006}. Finally,
the power spectrum based on X-ray data lacks a harmonic
corresponding to the $16.5-$s period. The phase
diagram clearly shows the contribution of one of the
spots to both the X-ray and UV emission from the
system, while the second spot contributes only in the
UV, with the X-ray intensity reaching its minimum at
this phase \cite{Choi-Dotani-2006}.

The final conclusion that the standard scenario with disk accretion onto the
surface of the white dwarf was not able to describe the mass transfer and
emission processes in the AE~Aquarii system was based on an analysis of
optical spectrograms, in particular, H$_\alpha$ Doppler tomograms. The
distance from the white dwarf to the point where the material approaches it
most closely exceeds the white dwarf radius by a factor of 30--70
\cite{Wynn1997, Ikhsanov2004}. It was also noted that the H$_\alpha$ Doppler
tomograms presented in \cite{Wynn1997, Welsh-etal-1998} showed no signs of the
presence of a developed Keplerian accretion disk in the system. On the
contrary, they correspond to the case when material flowing into the Roche
lobe of the white dwarf through the Lagrangian point L$_1$ leaves the system
after interacting with the magnetic field of the rapidly rotating white dwarf
(see \cite{Ikhsanov2012} and references therein).

The current model of AE~Aquarii based on the observations described above is
able to explain the rapid braking by the rotation of the white dwarf (in the
framework of the standard model for a radio pulsar) and the presence of hot
spots by the dissipation of the back current in its magnetosphere, and to
identify the additional extended radiation source with material flowing in the
Roche lobe of the white dwarf and leaving the system. At the same time, the
origin of the flaring activity of the object remains an open question in this
model. The lack of correlation of the flares with the orbital phase
\cite{Bruch-1991} and the high luminosity during the strongest flares
(comparable to the mean bolometric luminosity of the system
\cite{Beskrovnaya-etal-1996}) make it impossible to explain the flares in
terms of magnetic flaring activity of the red dwarf. It is  more plausible
that the source of the flares observed in the optical, UV, and X-ray is inside
the Roche lobe of the white dwarf. An increase in the effective area (volume)
of the extended source \cite{de-Martino-etal-1995, Beskrovnaya-etal-1996} and
a considerable broadening of the emission line wings
\cite{Reinsch-Beuermann-1994, Welsh-etal-1995} during flares support this
hypothesis. In this case, the brightness variations  reflect the
non-stationary character of the energy and angular momentum exchange between
the white dwarf and material flowing between the system components.

Until recently, the interaction between the magnetic field of the white dwarf
and the material flowing in its Roche lobe was studied using several
simplifying assumptions. In particular, the stream of matter was modeled as a
superposition of diamagnetic blobs, treated like test particles in the
computations. Deformation and heating of the blobs were neglected. The rate of
change of the kinetic energy of the blobs moving relative to the magnetic
field  of the white dwarf was normalized to its maximum value, and the
correction coefficient was estimated empirically, by comparing the
computational results to the velocity field derived from observed $H_{\alpha}$
Doppler tomograms  \cite{Wynn1997, Ikhsanov2004}. In this approach, the
velocity dispersion of the material outflow was interpreted in terms of the
differences in the masses and radii of the test particles, leading to
differences in the trajectories of their motion in the Roche lobe of the white
dwarf. The cause for  fragmentation of the initial stream into plasma
condensations was not considered.

The results of computations of mass transfer in low-mass binaries performed
using a full set of MHD equations confirm the possibility of a scenario where
a stream of matter leaves the system after interaction with the white dwarf
magnetosphere
 \cite{zbSMF, zbbUFN2012} (see also \cite{mcb-book}). Such objects have
acquired the name ``super-propellers''\footnote{Note that the term
``super-propeller'' is sometimes used in a somewhat different sense. For
example, the term ``superpropeller mode'' is applied to the propeller mode
under the conditions of super-critical accretion in \cite{Lipunov1987}.}
\cite{zbbUFN2012}. These magnetic white dwarfs rotate so rapidly that the
linear velocities at their magnetosphere boundaries considerably exceed the
Keplerian velocities. It was also noted that the process of interaction
between a superpropeller and the material surrounding its magnetosphere is
unstable, and its efficiency  depends significantly on the physical parameters
of matter located at the magnetospheric boundary. This provides a wide variety
of possible scenarios for mass transfer in the system (from the formation of a
transient disk to the complete suppression of mass exchange between the
components via the Lagrangian point L$_1$), including periodic replacement of
one flow type with another on the dynamical time scale.

In this study, we have analyzed possibilities for describing the plasma flows
in AE Aquarii in the superpropeller model. In Section\,2, we analyze the
efficiency of the interaction between the magnetosphere of the white dwarf and
the material flowing into its Roche lobe from L$_1$, taking into account the
conductivity of this stream. Section\,3 describes the numerical model used.
The results of our numerical computations of the mass transfer picture
presented in Section\,4 indicate possible formation of a transient disk in the
system, with a typical life time up to several hours. Observable
characteristics of the source within this scenario are discussed in
Section\,5.

\section{Interaction of the magnetic field
with the stream of plasma}

We consider a plasma stream flowing in the magnetic field of a rapidly
rotating white dwarf. Mass transfer between the components takes place in
semidetached binaries, leading to the formation of a gaseous stream which
leaves the donor star envelope through the inner Lagrangian point L$_1$,
enters the Roche lobe of the accreting star and moves  along almost a
ballistic trajectory.

The interaction between the stream material and the magnetic field of the
white dwarf can be fairly complex. We performed a detailed analysis of
possible physical mechanisms which could contribute to changing the efficiency
of the transfer of additional angular momentum from the rotating magnetosphere
to the material. In particular, we considered \cite{Isakova2014} the partial
ionization of plasma in the stream, the pressure of magnetodipole radiation or
a relativistic stellar wind from the rotating white dwarf, the relativistic
retard of the magnetic field lines, and incomplete penetration of the magnetic
field into  plasma. We have concluded that the latter effect probably plays
the most important role in the case of  AE~Aquarii system. It is accordingly
this effect that we considered in our numerical modeling and used  to explain
the observational manifestations of processes in AE~Aquarii.

Let us briefly elucidate the essence of this effect. The plasma stream has a
diamagnetic effect. This diamagnetism is related to the following known
property of plasma (see, for instance, \cite{Frank-Kamenetsky, Chen}). When a
clump of plasma enters an external magnetic field, currents are generated on
its surface, creating their own magnetic field. The current generation in this
case is such that the magnetic field induced in the plasma clump cancels out
the external magnetic field almost completely (apart from a thin surface
layer). The external field penetrates gradually, due to diffusion processes.

Initially, only the magnetic field which was present in the secondary envelope
is present in the stream. This field is determined by the magnetic field of
the white dwarf which has penetrated into this region, as well as by the
intrinsic magnetic field of the donor star \cite{Meintjes2004}. Due to
freezing of the field into the plasma, this magnetic field is transported into
the Roche lobe together with the   stream of plasma. However, this magnetic
field is much weaker than the magnetic field in the white dwarf magnetosphere.

Let us calculate the force of the interaction between
the plasma stream and the external magnetic
field. When a conducting body moves in a rarefied
magnetized plasma, electric fields and currents are
induced on the surface of the body. This results in
electromagnetic braking of the body, and the generation
of Alfv\'en and magnetosonic waves. This
type of interaction of conducting bodies during their
motion in a rarefied magnetized plasma is called an
induction interaction \cite{Drell1965,
Gurevich1978, Rafikov1999}.

We can apply these considerations to our case. Under our assumptions, the
total force acting on the stream is
\begin{equation}\label{eq-F1}
 {\vec F}_* = \frac{1}{c}
 \int\limits_{\delta V}
 \left( {\vec j} \times {\vec B} \right)
 dV,
\end{equation}
where $c$ is the speed of light, and ${\vec j}$ is the induced current
density in a surface layer of the stream occupying
a volume $\delta V$. The current density is determined by
the induced magnetic field arising due to the motion
of the gas stream in the external field:
\begin{equation}\label{eq-jE}
 {\vec j} = \hat{\sigma} \cdot {\vec E}, \quad
 {\vec E} = \frac{1}{c}
 \left( {\vec u} \times {\vec B} \right),
\end{equation}
where $\hat{\sigma}$ is the conductivity tensor of the stream material. For
simplicity, we neglected magnetization and assumed that the conductivity could
be described with a constant scalar coefficient $\sigma$ (as in MHD).
Substituting expressions \eqref{eq-jE} into \eqref{eq-F1}, we obtain the
estimate
\begin{equation}\label{eq-F2}
 {\vec F}_* = -\frac{\sigma}{c^2}
 \delta V B^2 {\vec u}_{\perp},
\end{equation}
where ${\vec u}_{\perp}$ is the component of the velocity ${\vec u}$ perpendicular
to the magnetic field lines. The corresponding
specific force is
\begin{equation}\label{eq-F3}
 {\vec f}_* = -\frac{\sigma B^2}{c^2 \rho}
 \frac{\delta V}{V} {\vec u}_{\perp},
\end{equation}
where $\rho$ is the density of the material in the stream.

We denote the diffusion coefficient of the magnetic field in the stream plasma
as $\eta = c^2/(4\pi\sigma)$. Then, we obtain the final expressions in the
form
\begin{equation}\label{eq-F4}
 {\vec f}_* = -\frac{{\vec u}_{\perp}}{t_w}, \quad
 t_w = \frac{4\pi\rho\eta}{B^2} \frac{V}{\delta V}.
\end{equation}
Note that, compared to the formula for the relaxation time $t_w$ used in our
earlier computations (e.g., \cite{zbSMF, zbbUFN2012}), Eq. \eqref{eq-F4}
contains an additional coefficient, $V/\delta V$, due to the presence of a
diffusion layer. In the case of complete penetration of the magnetic field
into the stream plasma, when the volume of the diffusion layer $\delta V$
coincides with the volume of part of   the stream $V$, Eq. \eqref{eq-F4} for
the relaxation time becomes simpler, and acquires the form
\begin{equation}\label{eq-t1}
 t_w = \frac{4\pi\rho\eta}{B^2}.
\end{equation}
In the opposite limiting case, when the diffusion layer
depth is $\delta
\ll R$ (where $R$ is the scale size of a plasma
clump), the volume ratio is $\delta V / V = \kappa \delta / R$, so that
\begin{equation}\label{eq-t2}
 t_w = \frac{4\pi\rho\eta}{B^2} \frac{R}{\kappa \delta}.
\end{equation}
The coefficient $\kappa$ is determined by the stream geometry.
For example, $\kappa = 3$ for a spherically symmetric
plasma clump.

Similar expressions for the interaction force between the gas stream and the
external magnetic field can  be obtained using the formalism of wave MHD
turbulence. This approach can apparently be applied in the case of a very
strong external magnetic field \cite{zbSMF}. The reason is that the plasma
dynamics in a strong external magnetic field is characterized by a relatively
slow mean motion of the particles along the magnetic field lines, a drift
across the field lines, and the propagation of Alfv\'en and magnetosonic waves
with velocities which are very high compared to their background. The
structure of such a flow can be described using an averaged representation,
considering the influence of the rapid pulsation by analogy to the wave MHD
turbulence. To describe the slow proper motion of the plasma, it is necessary
to isolate the rapidly propagating fluctuations and apply some averaging
procedure to the ensemble of wave pulsations. We will derive the main
expressions, following \cite{zbbUFN2012}, taking into account necessary
corrections.

Consider the expression
\begin{equation}\label{eq-ohm}
 {\vec E} + \frac{1}{c}
 \left( {\vec v} \times {\vec B} \right) =
 \frac{{\vec j}}{\sigma},
\end{equation}
which  describes the Ohm's law  for  plasma in the MHD approximation
\cite{Landau8}. Here, ${\vec B} = {\vec B}_{*} + {\vec b}$ is the total
magnetic field induction, which is the sum of the unperturbed external
magnetic field, ${\vec B}_{*}$, and the intrinsic field of the plasma, ${\vec
b}$. We emphasize that Eq. \eqref{eq-ohm} should be used only in the diffusion
layer into which the external magnetic field penetrates. No induction currents
appear in the remaining volume of the stream.

Let us write all dynamical quantities as sums of
their mean values and fluctuations, for example,
${\vec b} = \aver{{\vec b}} + \delta{\vec b}$. Averaging \eqref{eq-ohm}, we find
\begin{equation}\label{eq-ohm2}
 c\aver{{\vec E}} +
 \aver{{\vec v}} \times \aver{{\vec b}} +
 \aver{{\vec v}} \times {\vec B}_{*} +
 \aver{\delta{\vec v} \times \delta{\vec b}} =
 \frac{c}{\sigma} \aver{{\vec j}}.
\end{equation}
The last term on the left-hand side can be estimated
using the following equation, often used in dynamo
theory (e.g., \cite{Parker1982, Ruzmaikin1988}):
\begin{equation}\label{eq-dyn}
 \aver{\delta{\vec v} \times \delta{\vec b}} =
 \alpha \aver{{\vec b}} -
 \eta_w \rotor \aver{{\vec b}},
\end{equation}
where $\alpha$ is determined by the mean helicity of the
flow, and $\eta_w$ is the diffusion coefficient for the diffusion
of the mean magnetic field due to wave MHD
turbulence. We can neglect the first term (the $\alpha$
effect) because it describes the relatively weak and
slow generation of the mean magnetic field in the
accretion disk.

Averaging the Maxwell's equation,
\begin{equation}\label{eq-m1}
 \rotor {\vec B} = \frac{4\pi}{c} {\vec j},
\end{equation}
we find
\begin{equation}\label{eq-j1}
 \aver{{\vec j}} = \frac{c}{4\pi} \rotor \aver{{\vec b}}.
\end{equation}
Substituting \eqref{eq-dyn} and\eqref{eq-j1} into \eqref{eq-ohm2} yields
\begin{equation}\label{eq-ohm3}
 c\aver{{\vec E}} +
 \aver{{\vec v}} \times \aver{{\vec b}} +
 \aver{{\vec v}} \times {\vec B}_{*} -
 \eta_{w} \rotor \aver{{\vec b}} =
 \eta_{\text{OD}} \rotor \aver{{\vec b}},
\end{equation}
where $\eta_{\text{OD}} = \sfr{c^2}{(4\pi\sigma)}$~ is the ohmic diffusion
coefficient of the magnetic field. Since, as a rule, $\eta_w \gg
\eta_{\text{OD}}$, we can neglect the right-hand side. Moreover, in the case
of a strong magnetic field, we can neglect the second  term (compared to the
third).

Then, averaging the Maxwell's equation describing the law of electromagnetic
induction,
\begin{equation}\label{eq-m2}
 \rotor {\vec E} = -\frac{1}{c} \pdiff{{\vec B}}{t},
\end{equation}
we find
\begin{equation}\label{eq-E1}
 c \rotor \aver{{\vec E}} =
 -\pdiff{\aver{{\vec b}}}{t} -
 \rotor \left( {\vec v}_{*} \times {\vec B}_{*} \right),
\end{equation}
where ${\vec v}_*$ is the velocity of the magnetic field lines.
The first term on the right-hand side of this equation
is due to variations of the mean magnetic field, $\aver{{\vec b}}$,
on the dynamical time scale. This term corresponds
to the second term in \eqref{eq-ohm3} to order of magnitude.
Thus, the mean electric field strength can be estimated from
the expression
\begin{equation}\label{eq-E2}
 c \aver{{\vec E}} = -{\vec v}_{*} \times {\vec B}_{*}.
\end{equation}
Leaving the dominating terms in \eqref{eq-ohm3}, we arrive at the
relation
\begin{equation}\label{eq-ohm4}
 \eta_{w} \rotor \aver{{\vec b}} = {\vec u} \times {\vec B}_{*},
\end{equation}
where ${\vec u} = \aver{{\vec v}} - {\vec v}_{*}$ is the relative velocity of the
stream and magnetic field lines.

The obtained expression can be used to calculate
the average electromagnetic force in the equation
of motion. Neglecting density fluctuations, wave
magnetic pressure, and tension, we find for the full
electromagnetic force acting on the gas stream
\begin{equation}\label{eq-F10}
 {\vec F}_\text{em} =
 -\frac{1}{4\pi} \int\limits_V
 \aver{{\vec B} \times \rotor {\vec B}}\, dV =
 -\frac{1}{4\pi} \int\limits_V
 \aver{{\vec b}} \times \rotor \aver{{\vec b}}\, dV
 -\frac{1}{4\pi} \int\limits_{\delta V}
 {\vec B}_{*} \times \rotor \aver{{\vec b}}\, dV.
\end{equation}
The first term on the right-hand side describes the electromagnetic force due
to the intrinsic magnetic field of the plasma,  $\aver{{\vec b}}$. The second
term, describing the force from the external field, takes into account the
fact that the field ${\vec B}_*$ penetrates the plasma only in the diffusion
layer. We calculated this term using \eqref{eq-ohm4}. We have
\begin{equation}\label{eq-F11}
 {\vec B}_{*} \times \rotor \aver{{\vec b}} =
 \frac{B_*^2}{\eta_w}\, {\vec u}_{\perp},
\end{equation}
where the subscript $\perp$ denotes velocity components
perpendicular to the magnetic field ${\vec B}_{*}$.

We have obtained the following expression for the corresponding specific
force:
\begin{equation}\label{eq-F12}
 {\vec f}_* = -\frac{{\vec u}_{\perp}}{t_w}, \quad
 t_w = \frac{4\pi\rho\eta_w}{B^2} \frac{V}{\delta V}.
\end{equation}
In contrast to \eqref{eq-F4}, the meaning of the diffusion coefficient is
different here. However, if we do not go into the nature of the diffusion of
the magnetic field, these formulae appear equivalent. Moreover, all the above
calculations can be considered a modified derivation of the expression for the
force ${\vec f}_*$ in \eqref{eq-F4} in the framework of the induction
mechanism, taking into account the effects of the wave MHD turbulence which
appears in the case of a very strong external field.

\section{Description of the model}

We describe the flow structure in a close binary system using a non-inertial
reference frame rotating with the orbital angular velocity $\Omega = 2\pi /
P_\text{orb}$ relative to its center of mass. In this reference frame, we
choose a Cartesian coordinate system ($x$, $y$, $z$), with its origin
coincident with the center of the degenerate component, ${\vec r}_\text{a} =
0$. The center of mass of the donor star is located at the distance $A$ from
the origin along the $x$ axis, ${\vec r}_\text{d} = (-A, 0, 0)$. The $z$ axis
is along the rotation axis of the system, ${\vec \Omega} = (0, 0, \Omega)$. In
this reference frame, the rotation of the white dwarf is described by the
angular velocity vector ${\vec \Omega}_\text{a}$. In our model, we consider
the case when the rotation axis of the white dwarf coincides with the binary
rotation axis. Therefore\footnote{$P_\text{beat} = 2\pi/\Omega_\text{a}$ is
the beat period between the rotation period of the white dwarf,
$P_\text{spin}$, and the orbital period of the binary, $P_\text{orb}$. Because
of the very rapid rotation of the white dwarf in the AE~Aquarii system, the
beat period is $P_\text{beat} \approx P_\text{spin}$.},
\begin{equation}\label{eq-Omega_a}
 \Omega_\text{a} = \left(
 \frac{P_\text{orb}}{P_\text{spin}} - 1
 \right) \Omega,
\end{equation}
giving $\Omega_\text{a} = 1091.5\, \Omega$ for the AE~Aquarii system.

We  assume that the magnetic field of the white dwarf , ${\vec B}_{*}$, is
dipolar, and is described by the dipole moment vector ${\vec \mu}$. In the
chosen coordinates, ${\vec \mu}$ has the components $\mu_x = \mu \sin\theta
\cos\phi$, $\mu_y = \mu \sin\theta \sin\phi$, $\mu_z = \mu \cos\theta$, where
$\mu$ is the magnitude of the vector ${\vec \mu}$, $\theta$ is the inclination
of ${\vec \mu}$ to the $z$ axis, and $\phi$ is the angle between the $x$ axis
and the projection of ${\vec \mu}$ onto the $xy$ plane. Taking into account
the rotation of the white dwarf, the angle $\phi$ is time-dependent, so that
the magnetic field ${\vec B}_{*}$ is not stationary. In the approximation of a
quasi-stationary electromagnetic field, we can write: $\phi = \Omega_\text{a}
t + \phi_0$, where $\phi_0$ is the initial value of this angle. Note that, in
a quasi-stationary approximation, the magnetic field, ${\vec B}_{*}$, is a
potential field, $\rotor {\vec B}_{*} = 0$. This enables us to partially
exclude it from the equations describing the structure of the MHD flows in the
close binary system.

Taking into account the influence of the magnetic field, the flows of matter
in the close binary system can be described using the following system of
equations \cite{zbbUFN2012}:
\begin{equation}\label{eq-rho1}
 \pdiff{\rho}{t} + \nabla \cdot \left( \rho {\vec v} \right) = 0,
\end{equation}
\begin{equation}\label{eq-v1}
 \pdiff{{\vec v}}{t} + \left( {\vec v} \cdot \nabla \right) {\vec v} =
 -\frac{\nabla P}{\rho} -
 \frac{{\vec b} \times \rotor {\vec b}}{4 \pi \rho} -
 \nabla \Phi +
 2 \left( {\vec v} \times {\vec \Omega} \right) -
 \frac{\left( {\vec v} - {\vec v}_{*} \right)_{\perp}}{t_w},
\end{equation}
\begin{equation}\label{eq-b1}
 \pdiff{{\vec b}}{t} = \rotor \left[ {\vec v} \times {\vec b} +
 \left( {\vec v} - {\vec v}_{*} \right) \times {\vec B}_{*} -
 \eta\, \rotor {\vec b} \right],
\end{equation}
\begin{equation}\label{eq-s1}
 \rho T \left[ \pdiff{s}{t} +
 \left( {\vec v} \cdot \nabla \right) s \right] =
 n^2 \left( \Gamma - \Lambda \right) +
 \frac{\eta}{4 \pi} \left( \rotor {\vec b} \right)^2.
\end{equation}
Here, $\rho$ is the density in the stream, ${\vec v}$ is its velocity, $P$ is
the pressure, $s$ is the entropy per unit mass, $n = \rho/m_p$~ is the number
density, $m_p$ is the proton mass, and $\eta$ is the coefficient of magnetic
viscosity. The equation for the entropy (27) takes into account radiative
heating and cooling, as well as heating of material due to current
dissipation. The dependencies of the radiative heating function
 $\Gamma$ and the cooling function $\Lambda$ on the temperature $T$
have fairly complex forms \cite{Cox1971, Dalgarno1972, Raymond1976,
Spitzer1981}. Therefore, in our numerical model, we used linear approximations
for these functions in the vicinity of the equilibrium temperature, 8\,782~K
\cite{Bisikalo2003, zb2009, ZhilkinMM2010, mcb-book}, which corresponds to the
effective temperatures of all three sources of thermal radiation in the
AE~Aquarii system. The term $2 ({\vec v} \times {\vec \Omega})$ in the
equation of motion \eqref{eq-v1} describes the Coriolis force. The density,
entropy, and pressure are related through the ideal gas law: $s = c_V \ln(P /
\rho^{\gamma})$, where $c_V$ is the specific heat capacity of the gas at
constant volume, and the adiabatic index is $\gamma = 5/3$.

The numerical model takes into account the effects of magnetic field diffusion
[see~\eqref{eq-b1}] due to magnetic reconnection and current dissipation in
turbulent vortexes \cite{BisnovatyiKogan1976}, magnetic buoyancy
\cite{Campbell1997}, and wave MHD turbulence \cite{zbSMF}. The first two
effects play important role in the accretion disks of intermediate polars. The
third effect dominates in the region of the magnetosphere and in the accretion
streams of polars. The total magnetic viscosity coefficient $\eta$, determined
by all these effects, depends on the magnetic field in the plasma, ${\vec b}$.
Thus, in general, the character of the magnetic field diffusion is non-linear.

The last term in the equation of motion \eqref{eq-v1} describes the force
exerted by the magnetic field of the white dwarf  on the plasma. This force
determines the variations of the plasma velocity perpendicular to the magnetic
field, ${\vec v}_{\perp}$. The vector
\begin{equation}\label{eq-vs}
 {\vec v}_{*} = {\vec \Omega}_\text{a} \times
 \left( {\vec r} - {\vec r}_\text{a} \right)
\end{equation}
describes the velocity of the magnetic field
lines of the white dwarf. The decay time scale for the transverse
velocity $t_w$ is determined by wave dissipation of the
magnetic field in the diffusion layer.

We applied the 3D parallel numerical code \cite{zb2009, ZhilkinMM2010, zbSMF,
mcb-book} in our MHD numerical simulation. This code is based on the
Godunov-type difference scheme with the high order of accuracy. The
computations used a geometrically adaptive grid \cite{ZhilkinMM2010}, which
concentrated towards the equatorial plane and the surface of the white dwarf.
This made it possible to considerably improve the resolution of vertical
structures in the accretion stream and in the region of the white dwarf
magnetosphere.

\section{Discussion}

We have adopted the following model parameters for the close binary system
AE~Aquarii \cite{Ikhsanov2012}. The donor star (a red dwarf) has the mass
$M_\text{d} = 0.91 M_{\odot}$ and the effective temperature of $4\,000$~K. The
mass of the white dwarf is $M_\text{a} = 1.2 M_{\odot}$, and its effective
temperature is about $13\,000$~K. The component separation is $A = 3.02
R_{\odot}$. We assumed a magnetic field on the surface of the white dwarf
$B_\text{a} = 50$~MG.

We have used the following boundary conditions in our computations. At the
inner Lagrangian point L$_1$, we take the gas velocity to be equal to the
local sound speed,  $c_s = 7.4$~km/s, corresponding to the effective
temperature of the donor star of 4\,000~K. The gas density is $\rho(L_1) = 4.7
\times 10^{-8}$~g/cm$^3$, corresponding to the mass exchange rate $\dot{M} =
10^{-9} M_{\odot}$/year. We have imposed constant boundary conditions at the
other boundaries of the computation region. We obtained solutions in the
region ($-0.53A \le x \le 0.53A, -0.53A \le y \le 0.53A, -0.26A \le z \le
0.26A$), on a grid of $128 \times 128 \times 64$ cells.

To describe the flaring activity, we have analyzed the situation when the flow
of plasma in the Roche lobe of the white dwarf  occurs successively in two
modes: laminar and turbulent. In the laminar mode, the depth of the boundary
layer in the region where material interacts with the magnetic field of the
white dwarf is determined by microscopic processes, and turns out to be
relatively small. Under these conditions, the influence of the magnetic field
on the stream is relatively weak, and the parameters of its motion do not
differ strongly from those computed in a gas-dynamical approximation. In the
turbulent mode, the depth of the diffusion layer sharply increases. In this
case, the magnetic field penetrates deeply into the plasma of the stream and
has a significant influence on its motion.

In the laminar mode, the stream bends around the magnetosphere of the white
dwarf and forms a transient disk (ring). The transition of the disk to the
turbulent mode occurs during its interaction with the  stream arriving from
the donor or/and the development of a (Kelvin--Helmholtz type) instability at
the inner boundary of the disk. In the turbulent mode, the material of the
transient disk quickly mixes with the magnetic field of the white dwarf, and
is accelerated in the tangential direction and ejected from the Roche lobe.
Further, the flow of matter which continues to arrive from L$_1$ returns to
the laminar mode.

It is difficult to estimate the instantaneous depth
of the diffusion layer in the numerical model we used.
Instead, we assumed a corresponding magnetic viscosity
coefficient based on some phenomenological
considerations. This approach seems quite acceptable
for the purposes of this study.

\begin{figure}[ht!]
\centering
\includegraphics[width=0.7\textwidth,natwidth=322,natheight=211]{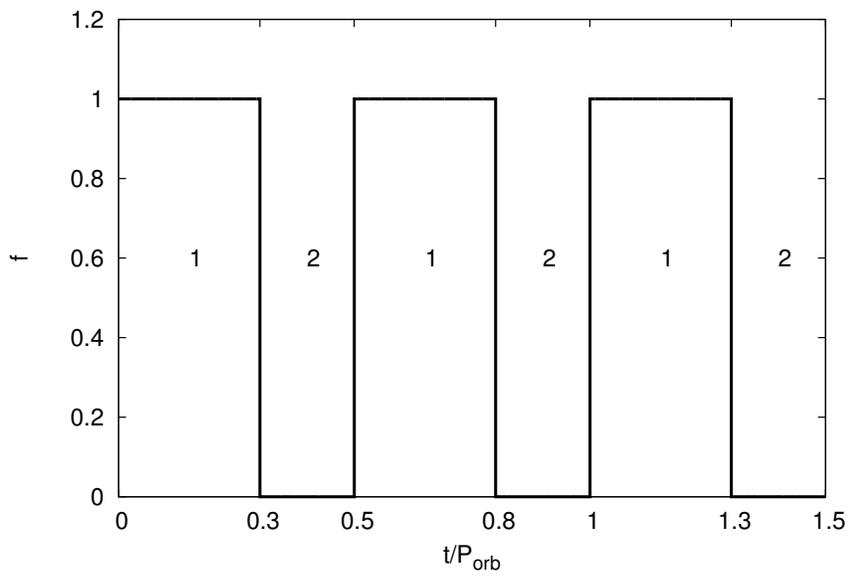}
\caption{ The function f(t). $1$ is the laminar phase, and $2$ is the
turbulent phase.} \label{fig-f}
\end{figure}

We have modeled the structure of the plasma flow in the Roche lobe of the
white dwarf, taking into account transitions between the laminar and turbulent
modes, using the numerical model described above, where the depth of the
diffusion layer is expressed as
\begin{equation}\label{eq-dV}
 \frac{\delta V}{V} = 1 - \left( 1 - e^{-k r} \right) f(t).
\end{equation}
Here $r$ is the distance from the rotation axis to a given point and $k$ is a
coefficient determining the characteristic size of the white dwarf
magnetosphere. The coefficient used in our computations was $k \approx 33/A$.
The function $f(t)$, displayed in Fig. \ref{fig-f}, is periodic, with the
period $P_\text{orb}/2$. In the initial phases of the orbital period ($0 \le t
\le 0.3 P_\text{orb}$), this function is equal to unity. In this case, the
depth of the diffusion layer \eqref{eq-dV} is virtually zero, with the
exception of the magnetosphere region close to the white dwarf. The plasma
flow in the Roche lobe is laminar at this stage. At later phases of the
orbital period ($0.3 P_\text{orb} \le t \le 0.5 P_\text{orb}$), $f(t) = 0$. In
this case, the depth of the diffusion layer \eqref{eq-dV} is the largest
($\delta V/V = 1$), with the result that the plasma flow in the Roche lobe
switches to the turbulent mode.

\begin{figure}[ht!]
\centering
\begin{tabular}{cc}
\hbox{\includegraphics[width=0.49\textwidth]{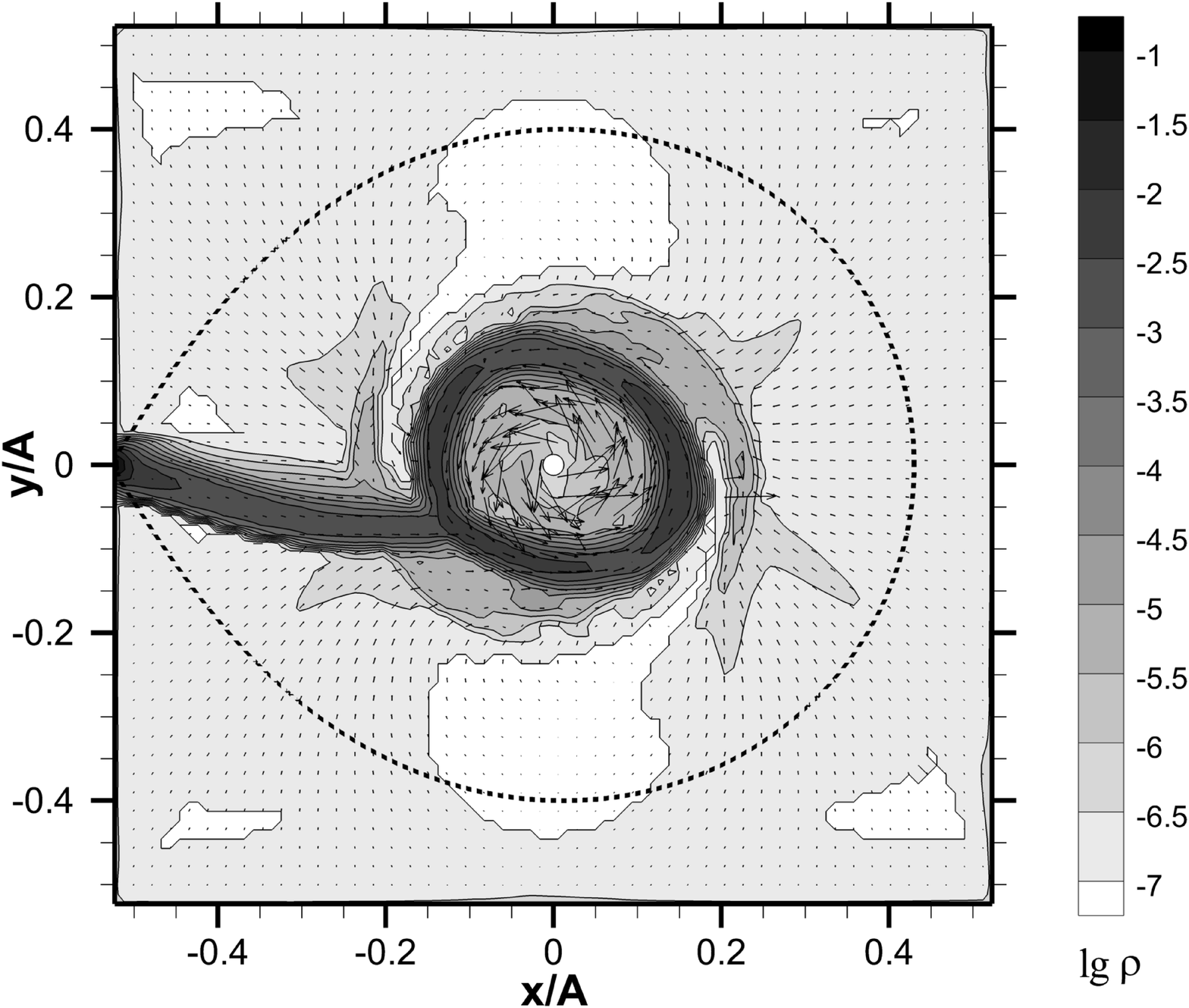}} &
\hbox{\includegraphics[width=0.49\textwidth]{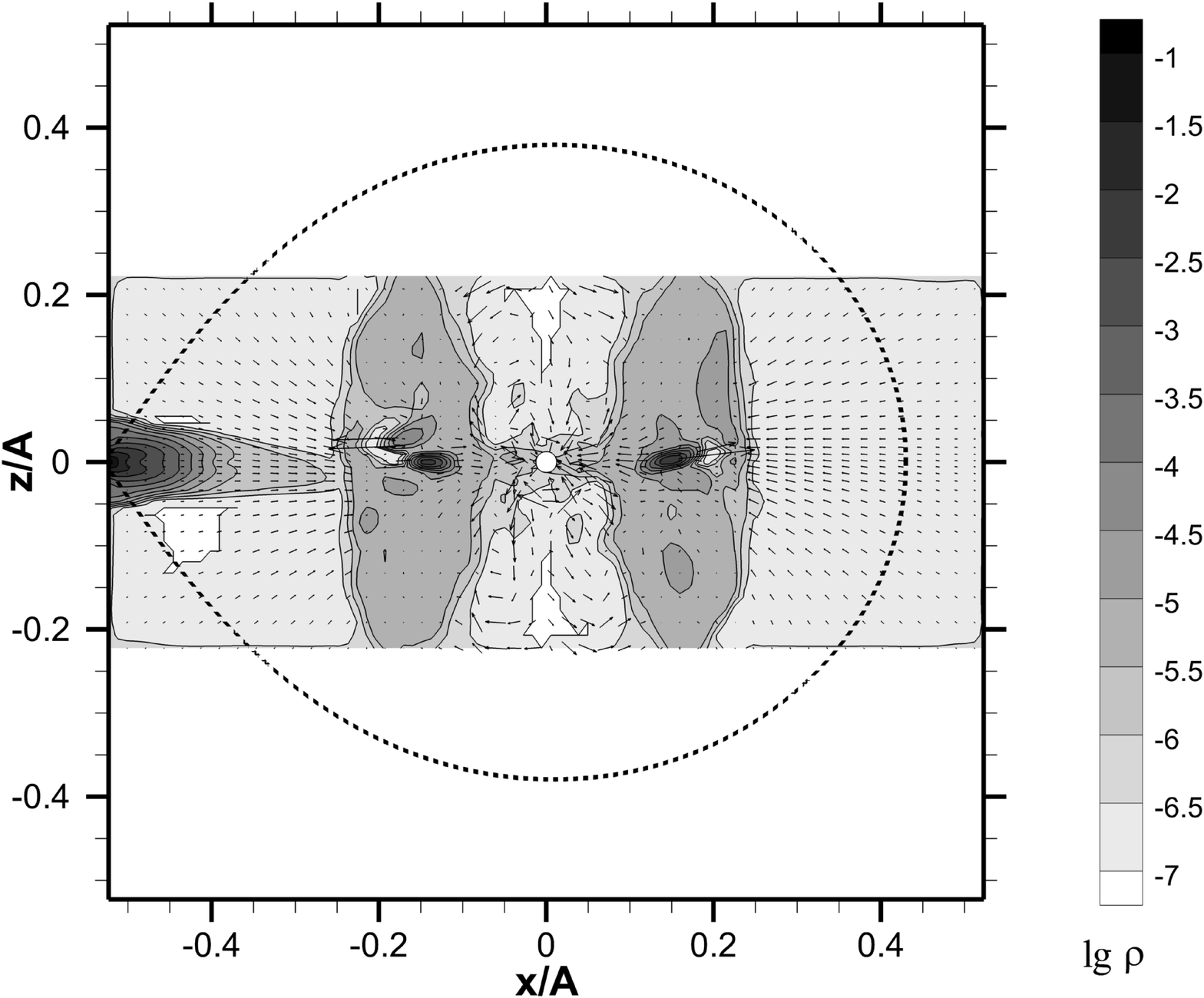}} \\
a) $xy$ & b) $xz$\\
\end{tabular}
\caption{The stationary laminar phase. The density and velocity distributions in the
equatorial (left) and vertical (right) planes of
the binary system.}
\label{fig1}
\end{figure}

\begin{figure}[ht!]
\centering
\begin{tabular}{cc}
\hbox{\includegraphics[width=0.49\textwidth]{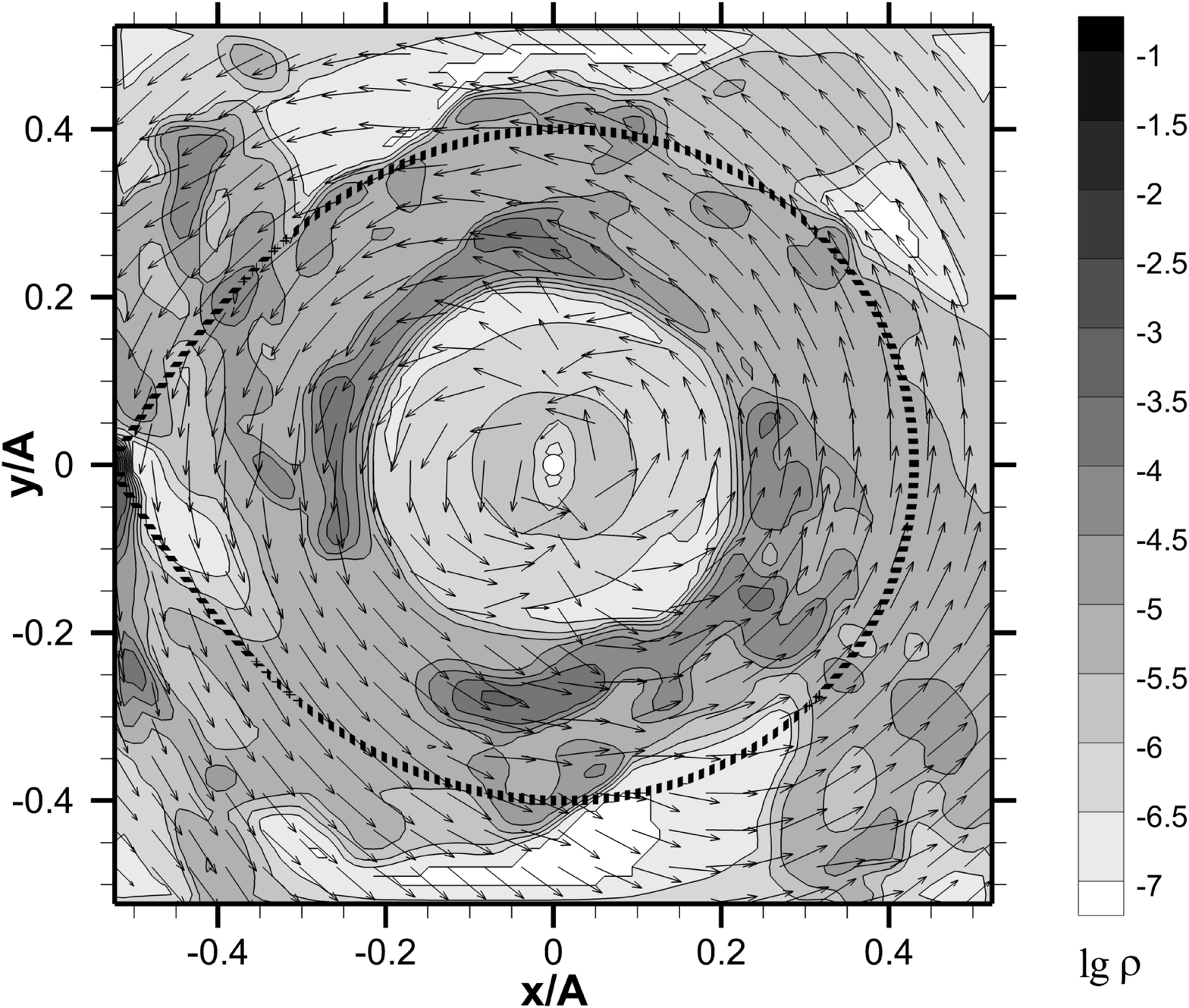}} &
\hbox{\includegraphics[width=0.49\textwidth]{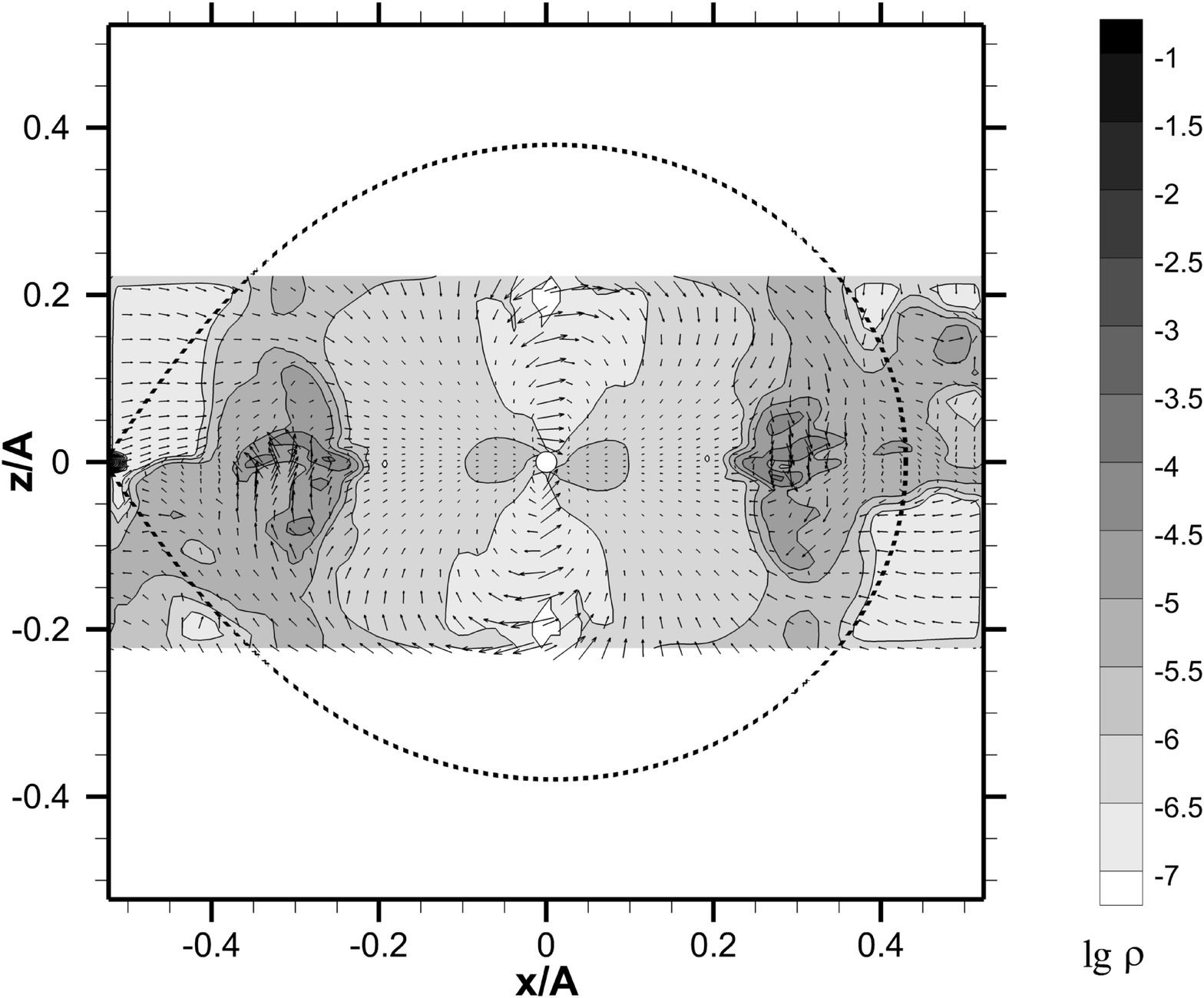}} \\
a) $xy$ & b) $xz$\\
\end{tabular}
\caption{The transition phase. The density and velocity distributions in the equatorial (left) and vertical (right) planes of the
binary system.}
\label{fig2}
\end{figure}

\begin{figure}[ht]
\centering
\begin{tabular}{cc}
\hbox{\includegraphics[width=0.49\textwidth]{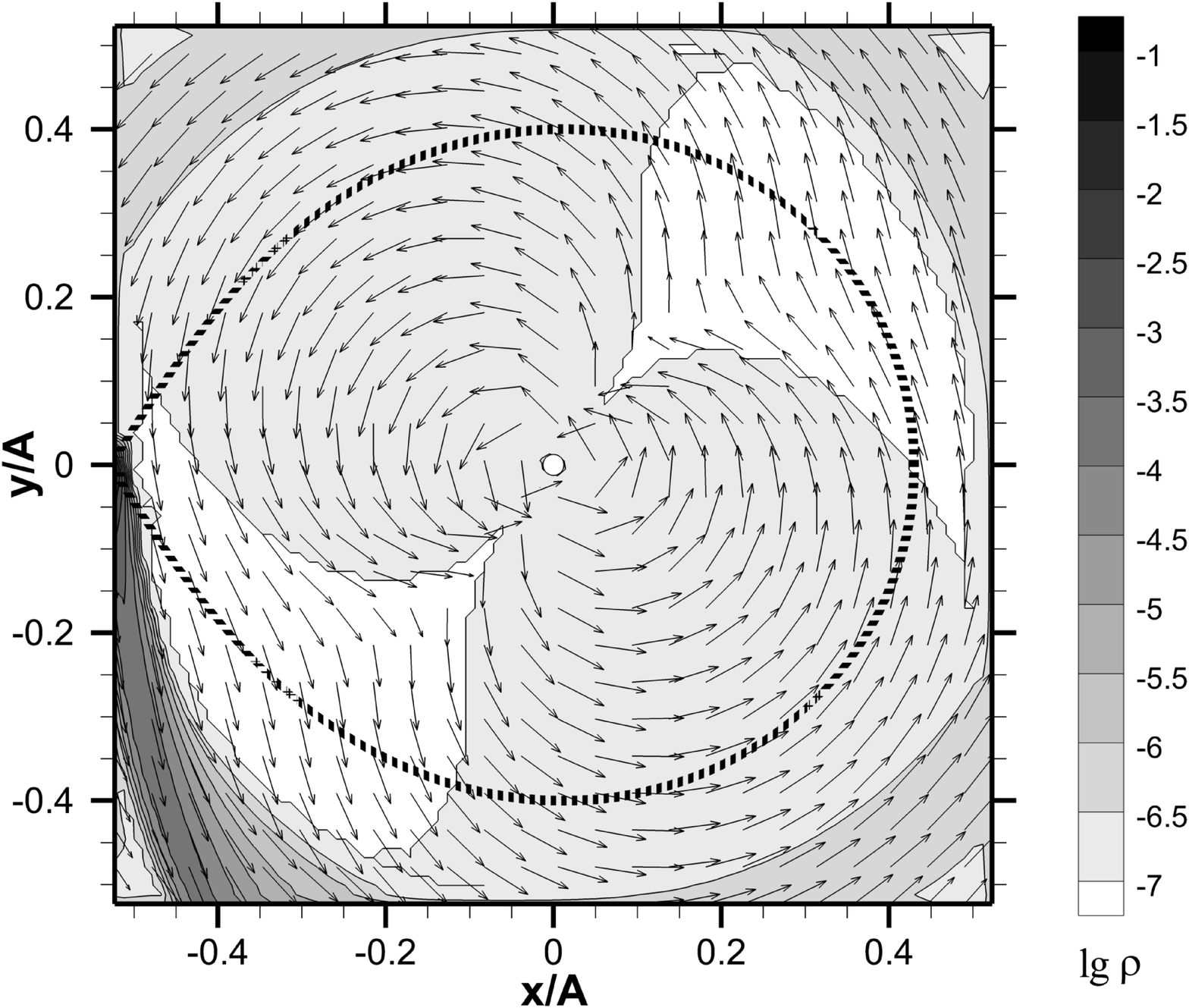}} &
\hbox{\includegraphics[width=0.49\textwidth]{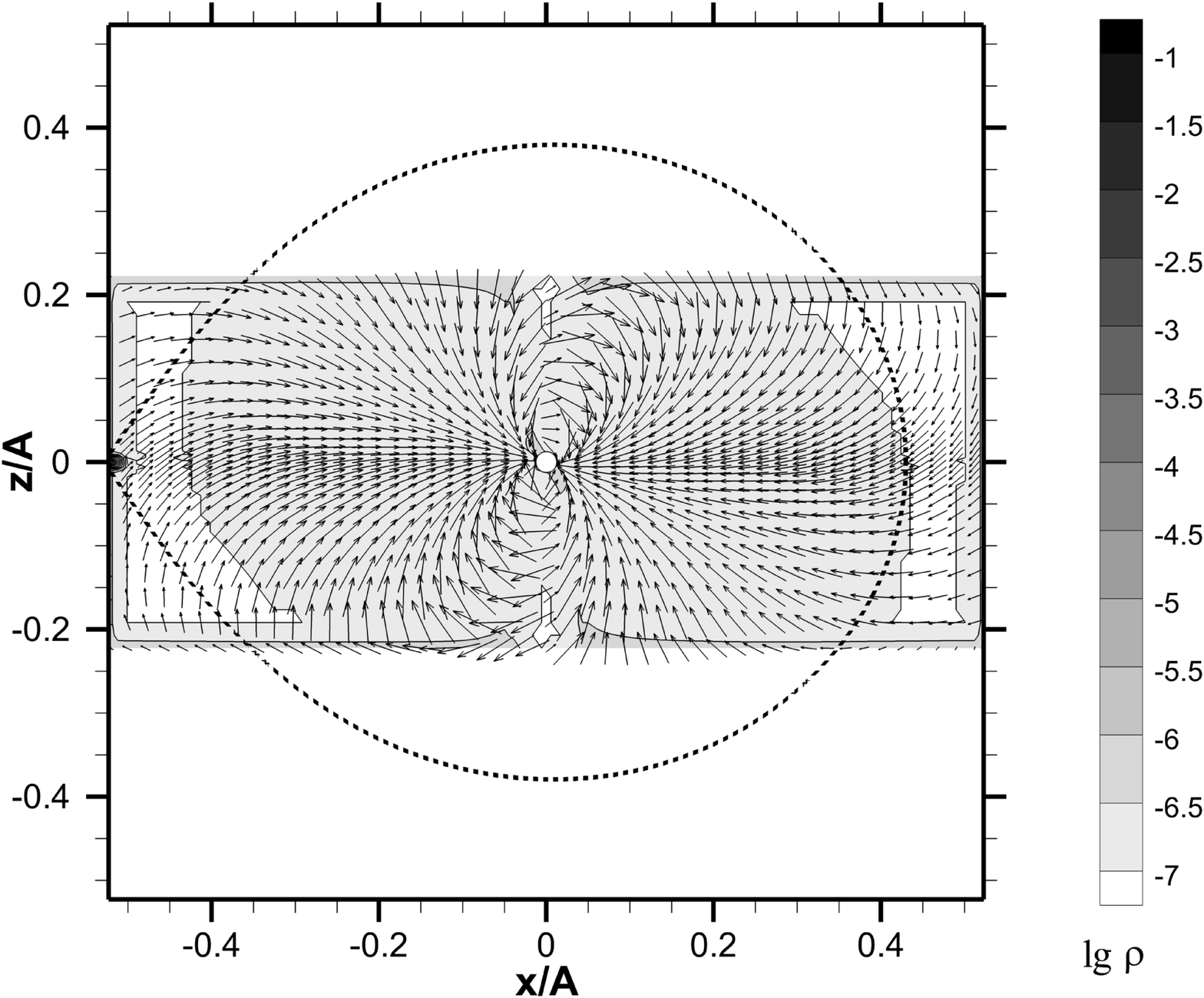}} \\
a) $xy$ & b) $xz$\\
\end{tabular}
\caption{The stationary turbulent phase. The density and velocity distributions in the equatorial (left) and vertical (right) planes
of the binary system.}
\label{fig3}
\end{figure}

The results of our numerical modeling of the AE Aquarii system in the proposed
scenario are presented in Figs. \ref{fig1}--\ref{fig3}. The gray scale
represents the distribution of the logarithm of the density in units of the
density at the inner Lagrangian point, $\rho(L_1) = 4.7 \times
10^{-8}~\text{g}/\text{cm}^3$, corresponding to the mass transfer rate
$\dot{M} = 10^{-9}~M_\odot/\text{year}$. The arrows show the velocity
distribution. The dotted curve corresponds to the boundary of the Roche lobe.
The white dwarf is shown as an open circle. In all these figures, the left
panel corresponds to the equatorial ($xy$) plane of the binary and the right
panel to the vertical ($xz$) plane.

In the quiescent phase (Fig. \ref{fig1}), the flow in the system is laminar.
The diffusion layer is thin in this case, and the magnetic field of the white
dwarf penetrates into the stream plasma only insignificantly. At the same
time, however, hot, rarefied plasma rotating with the magnetic field lines can
be present in the magnetosphere region. This forms a kind of corona around the
rapidly rotating white dwarf, and makes it more difficult for material to
penetrate this zone. In other regions, the magnetic field of the white dwarf
has virtually no influence on the plasma flow. We call this the stationary
laminar phase.

At this stage, the system forms a transient disk
with a typical radius of $0.1A$--$0.2A$ (left panel of
Fig. \ref{fig1}). The inner disk radius is determined by the
equilibrium condition at the boundary of the white dwarf
magnetosphere. The vertical structure of the
flow is displayed in the right panel of Fig. \ref{fig1}. The
white dwarf magnetosphere has the shape of a cylinder
with fairly broad and relatively dense walls. The
transient disk is located in the region of these walls,
in the equatorial plane of the binary.


After the end of the laminar phase, which lasts for $0.3 P_\text{orb} \approx
3$ hrs in our model, the numerical code switches to the computational regime
of the turbulent motion. The magnetic field diffusion coefficient increases
rapidly in the turbulent flow compared to the laminar flow, by several orders
of magnitude. The depth of the diffusion layer, where the plasma is frozen in
the magnetic field of the stream, increases considerably. The boundary of the
turbulence region propagates in the material with a velocity of order of the
velocity of magnetosonic waves. Thus, the entire flow region becomes turbulent
on a time scale shorter than the dynamical time scale. For some time, material
frozen in the magnetic field rotates rigidly, following the magnetic field
lines. The initial ring is spread into a disk, and material is gradually
carried beyond the limits of the Roche lobe (see Fig. \ref{fig2}). This flow
regime can be considered a transition phase. At this time, the surface area of
the disk increases, increasing its luminosity that is observationally
manifested  as a flare. The area of the expanding disk in Fig. \ref{fig2} is
approximately a factor of five to seven larger than the area of the transient
disk in Fig. \ref{fig1}. Note that the increase in the magnetic viscosity in
the transition to the turbulent mode should lead to an increase in current
dissipation, and thus to a temperature increase. This is an additional factor
increasing the luminosity. Thus, it is not difficult in our model to obtain a
luminosity of the system which is an order of magnitude higher during the
active phase than during the quiescence. This is precisely the rise of
luminosity
 which is observed during the active phase of AE~Aquarii.

The rate of material outflow beyond the limits of the system corresponds to
the dynamical time scale at the boundary of the Roche lobe of the white dwarf.
Matter arriving to the Roche lobe of the white dwarf through L$_1$ also leaves
the system during this phase, interacting with the material outflow and the
magnetic field of the white dwarf (see Fig. \ref{fig3}). This flow mode can be
considered a stationary turbulent phase, and corresponds to the usual
super-propeller state \cite{zbbUFN2012}. The duration of this stage was taken
to be $0.2 P_\text{orb}$ in our model. During all this time, the luminosity of
the system  decreases due to the decreasing radiation flux from the shell
ejected beyond the Roche lobe. Our earlier computations of such flows show
that the material outflow from L$_1$ forms a long trail which winds around the
binary in a spiral pattern, forming a common envelope. Similar results for
AE~Aquarii were also obtained in other studies using the Smoothed Particle
Hydrodynamics (SPH) method \cite{Wynn1997, Ikhsanov2004}.

\section{Conclusions}

The results of our computations show a wide variety of possible  flow regimes
in the Roche lobe of the rapidly rotating magnetic white dwarf in AE~Aquarii.
The flow pattern depends critically on the physical conditions in the material
interacting with the magnetic field, and can change from a transient Keplerian
disk (ring) to an intense, quasi-radial flow leaving the binary. Switching
between the flow modes occurs on the dynamical (free-fall) time scale, which
in AE~Aquarii ranges from several hundred seconds, in the region where
material approaches the white dwarf most closely, to several thousand seconds,
in the region of L$_1$. This corresponds to the characteristic time scale of
the flaring activity of the system. The duration of short flares, several
minutes, corresponds to the plasma turbulization time at the inner radius of
the transient disk. This results in plasma heating and acceleration, with
subsequent outflow from the system. The process of forming a new transient
disk takes about an hour. During this time, the system is in quiescence.

The stationary flow of the stream through the Roche lobe of the white dwarf,
without the formation of a Keplerian disk, discussed earlier in
\cite{Wynn1997, Ikhsanov2004}, is one particular solution of the problem we
have analyzed. This solution corresponds to the case of a turbulent stream
whose interaction with the magnetic field of the white dwarf is close to
maximally efficient. However, the question about the mechanism leading to the
turbulization of the stream in such a flow pattern remains open. There are no
collisions between clumps with different masses in the course of their motion
in the Roche lobe of the white dwarf \cite{Ikhsanov2004}. On the other hand,
the magnetic field at the boundary of this Roche lobe is relatively weak for
the parameters of interest for us, and it does not significantly influence the
 flow structure or its inner conditions. Finally, the relaxation time
scale for turbulent motions in the stream associated with Landau damping does
not exceed the dynamical time at the boundary of the Roche lobe of the white
dwarf. Therefore, the stream moving along a ballistic trajectory in the Roche
lobe of the white dwarf has enough time for the transition to the laminar flow
regime. This implies that the outflow of matter from the system can only take
place in the case of repeated turbulization of the stream in the region of its
strongest interaction with the magnetic field of the white dwarf, which is
confirmed by the results
of our computations.\\


\begin{theacknowledgments}
P.B.I., A.G.Z. and D.V.B. were supported by the Russian Science Foundation
(project 15-12-30038). N.R.I. and N.G.B. were supported by the Russian
Foundation for Basic Research (project 13-02-00077).
\end{theacknowledgments}


\begin{thebibliography}{99}

\bibitem{Warner}
B. Warner, \textit{Cataclysmic Variable Stars} (Cambridge: Cambridge Univ.
Press 2003).

\bibitem{Welsh-etal-1995}
W.F. Welsh, K. Horne, and R. Gomer, Monthly. Not. Roy. Astron. Soc. {\bf
275}, 649 (1995).

\bibitem{friedjung97}
 M. Friedjung, New Astron. {\bf 2}, 319 (1997).

\bibitem{Patterson-1979}
 J.\,Patterson, Astrophys. J. {\bf 234}, 978 (1979).

\bibitem{Reinsch-Beuermann-1994}
K. Reinsch and K. Beuermann, Astron. and Astrophys. {\bf 282}, 493
(1994).

\bibitem{Ikhsanov2012}
N. R. Ikhsanov and N. G. Beskrovnaya, Astron. Rep. {\bf 56}, 589 (2012).

\bibitem{Bastian-etal-1988}
T.S. Bastian, G.A. Dulk, and G. Chanmugam, Astrophys. J. {\bf 324}, 431 (1988).

\bibitem{Henize-1949}
K.G. Henize, Astrophys. J. {\bf 54}, 89 (1949).

\bibitem{Beskrovnaya-etal-1996}
N.G. Beskrovnaya, N.R. Ikhsanov, A. Bruch, and N.M. Shakhovskoy, Astron. Astrophys.
{\bf 307}, 840 (1996).

\bibitem{Bruch-1991}
A. Bruch, Astron. Astrophys. {\bf 251}, 59 (1991).

\bibitem{Eracleous-Horne-1996}
M. Eracleous and K. Horne, Astrophys. J. {\bf 471}, 427 (1996).

\bibitem{de-Jager-1994}
O.C. de~Jager, Astrophys. J. Suppl. Ser. {\bf 90}, 775 (1994).

\bibitem{Abada-Simon-etal-2005}
M. Abada-Simon, J. Casares, A. Evans, S. Eyres, R. Fender, S. Garrington, O. de~Jager,
N. Kuno, I.G. Martinez-Pais, D. de~Martino, H. Matsuo, M. Mouchet, G. Pooley, G. Ramsay,
A. Salama, and B. Schulz, Astron. Astrophys. {\bf 433}, 1063 (2005).

\bibitem{de-Jager-etal-1994}
O.C. de~Jager, P.J. Meintjes, D. O'Donoghue, and E.L. Robinson, Monthly. Not. Roy.
Astron. Soc. {\bf 267}, 577 (1994).

\bibitem{Welsh-1999}
W.F. Welsh, Proc. Annapolis Workshop on Magnetic Cataclysmic Variables,
eds. C.~Hellier and K.~Mukai, ASP Conference Series {\bf 157}, 357 (1999).

\bibitem{Ikhsanov-1998}
N.R. Ikhsanov, Astron. Astrophys. {\bf 338}, 521 (1998).

\bibitem{Ikhsanov-Biermann-2006}
N.R. Ikhsanov and P.L. Biermann, Astron. Astrophys. {\bf 445}, 305 (2006).

\bibitem{Eracleous-etal-1994}
M. Eracleous, K. Horne, E.L. Robinson, E.H. Zhang, T.R. Marsh, and J.H. Wood,
Astrophys. J. {\bf 433}, 313 (1994).

\bibitem{Osborne-etal-1995}
J.P. Osborne, K.L. Clayton, D. O'Donoghue, M. Eracleous, K. Horne, and A. Kanaan,
in Proc. Magnetic Cataclysmic Variables, Eds. D. Buckley and B. Warner, ASP
Conference Series {\bf 85}, 368 (1995).

\bibitem{Choi-etal-1999}
C.-S. Choi, T. Dotani, and P.C. Agrawal, Astrophys. J. {\bf 525}, 399 (1999).

\bibitem{Choi-Dotani-2006}
C.-S. Choi and T. Dotani, Astrophys. J. {\bf 646}, 1149 (2006).

\bibitem{Itoh-etal-2006}
K. Itoh, S. Okada, M. Ishida, and H. Kunieda, Astrophys. J. {\bf 639}, 397 (2006).

\bibitem{Wynn1997}
G.A. Wynn, A.R. King, and K. Horne, Monthly Notices Roy. Astron. Soc.
\textbf{286}, 436 (1997).

\bibitem{Ikhsanov2004}
N.R. Ikhsanov, V.V. Neustroev, and N.G. Beskrovnaya, Astron. and Astrophys.
\textbf{421}, 1131 (2004).

\bibitem{Welsh-etal-1998}
W.F. Welsh, K. Horne, and R. Gomer, Monthly. Not. Roy. Astron. Soc. {\bf 298}, 285
(1998).

\bibitem{de-Martino-etal-1995}
D. de~Martino, W. Wamsteker, and G. Bromage, in Proc. Magnetic Cataclysmic
Variables, Eds. D. Buckley and B. Warner, ASP Conference Series {\bf 85}, 388
(1995).

\bibitem{zbSMF}
A. G. Zhilkin and D. V. Bisikalo, Astron. Rep. \textbf{54}, 1063 (2010).

\bibitem{zbbUFN2012}
A. G. Zhilkin, D. V. Bisikalo, and A. A. Boyarchuk,
Phys. Usp. \textbf{55}, 115 (2012).

\bibitem{mcb-book}
D. V. Bisikalo, A. G. Zhilkin, and A. A. Boyarchuk,
\textit{Gas-Dynamics of Binary Stars} (Fizmatlit, Moscow,
2013) [in Russian].

\bibitem{Lipunov1987}
V. M. Lipunov, \textit{Astrophysics of Neutron Stars}
(Nauka, Moscow, 1987; Springer, Heidelberg, 1992).

\bibitem{Isakova2014}
P. Isakova, A. Zhilkin, and D. Bisikalo, EPJ Web of Conferences, \textbf{64},
id.03002 (2014).

\bibitem{Frank-Kamenetsky}
D. A. Frank-Kamenetskii, \textit{Lectures on Plasma
Physics} (Atomizdat, Moscow, 1968) [in Russian].

\bibitem{Chen}
F. F. Chen, \textit{Introduction to Plasma Physics and
Controlled Fusion} (Springer Science, New York,
1984; Moscow,Mir, 1987).

\bibitem{Meintjes2004}
P.J. Meintjes, Monthly. Not. Roy. Astron. Soc. {\bf 352}, 416 (2004).

\bibitem{Drell1965}
S.D. Drell, H.M. Foley, and M.A. Ruderman, J. Geophys. Res., \textbf{70}, 3131 (1965).

\bibitem{Gurevich1978}
A. V. Gurevich, A. L. Krylov, and E. N. Fedorov, Sov.
Phys. JETP \textbf{48}, 1074 (1978).

\bibitem{Rafikov1999}
R. R. Rafikov, A. V. Gurevich, and K. P. Zybin, J. Exp.
Theor. Phys. \textbf{88}, 297 (1999).

\bibitem{Landau8}
L. D. Landau and E.M. Lifshits, Course of Theoretical
Physics, Vol. 8: \textit{Electrodynamics of Continuous
Media} (Fizmatlit, Moscow, 2003; Pergamon, New
York, 1984).

\bibitem{Parker1982}
E. Parker, \textit{Cosmical Magnetic Fields} (Claredon,
Oxford, 1979).

\bibitem{Ruzmaikin1988}
A. A. Ruzmaikin, D. D. Sokolov, and A.M. Shukurov,
\textit{Magnetic Fields of Galaxies} (Nauka, Moscow,
1988; Kluwer, Dordrecht, 1988).

\bibitem{Cox1971}
D.P. Cox and E. Daltabuit, Astrophys. J. \textbf{167}, 113
(1971).

\bibitem{Dalgarno1972}
A. Dalgarno and R.A. McCray, ARA\&A, 375 (1972).

\bibitem{Raymond1976}
J.C. Raymond, D.P. Cox, and B.W. Smith, Astrophys. J. \textbf{204}, 290 (1976).

\bibitem{Spitzer1981}
L. Spitzer, \textit{Physical Processes in the Interstellar
Medium} (Wiley, New York, 1978; Mir, Moscow,

\bibitem{Bisikalo2003}
D.V.Bisikalo, A. A. Boyarchuk, P. V.Kaigorodov, and
O. A. Kuznetsov, Astron. Rep. \textbf{47}, 809 (2003).

\bibitem{zb2009}
A. G. Zhilkin and D. V. Bisikalo, Astron. Rep. \textbf{53}, 436
(2009).

\bibitem{ZhilkinMM2010}
A. G. Zhilkin, Mat. Model. \textbf{22}, No. 1, 110 (2010).


\bibitem{BisnovatyiKogan1976}
G.S. Bisnovatyi-Kogan and A.A. Ruzmaikin, Astrophys. Space. Sci. \textbf{42},
401 (1976).

\bibitem{Campbell1997}
C.G. Campbell, \textit{Magnetohydrodynamics in binary stars} (Dordrecht:
Kluwer Acad. Publishers 1997).

\end{thebibliography}
\end{document}